\documentclass[twocolumn,showpacs,preprintnumbers,amsmath,amssymb,prb]{revtex4}

\usepackage{graphicx}
\usepackage{dcolumn}
\usepackage{bm}
\usepackage{tabularx}
\usepackage{amsmath}

\begin{document}

\title{Specific heat of the iron-based high-$T_c$ superconductor SmO$_{1-x}$F$_x$FeAs}

\author{L. Ding,$^1$ C. He,$^1$ J. K. Dong,$^1$ T. Wu,$^2$ R. H. Liu,$^2$ X. H. Chen,$^2$ and S. Y. Li$^{1,*}$}

\affiliation{$^1$Department of Physics and Laboratory of Advanced Materials, Fudan University, Shanghai 200433, P. R. China\\
$^2$Hefei National Laboratory for Physical Science at Microscale and
Department of Physics, University of Science and Technology of
China, Hefei, Anhui 230026, P. R. China}

\date{\today}

\begin{abstract}
The specific heat $C(T)$ of new iron-based high-$T_c$ superconductor
SmO$_{1-x}$F$_x$FeAs ($0 \leq x \leq 0.2$) was systematically
studied. For undoped $x$ = 0 sample, a specific heat jump was
observed at 130 K. This is attributed to the structural or
spin-density-wave (SDW) transition, which also manifests on
resistivity as a rapid drop. However, this jump disappears with
slight F doping in $x$ = 0.05 sample, although the resistivity drop
still exists. The specific heat $C/T$ shows clear anomaly near $T_c$
for $x$ = 0.15 and 0.20 superconducting samples. Such anomaly has
been absent in LaO$_{1-x}$F$_x$FeAs. For the parent compound
SmOFeAs, $C(T)$ shows a sharp peak at 4.6 K, and with electron
doping in $x$ = 0.15 sample, this peak shifts to 3.7 K. It is
interpreted that such a sharp peak results from the
antiferromagnetic ordering of Sm$^{3+}$ ions in this system, which
mimics the electron-doped high-$T_c$ cuprate
Sm$_{2-x}$Ce$_x$CuO$_{4-\delta}$.

\end{abstract}


\maketitle

The recent discovery of superconductivity at $T_c$ = 26 K in
iron-based LaO$_{1-x}$F$_x$FeAs ($x$ = 0.05 - 0.12) \cite{Kamihara}
has attracted great attention. Following this initial work, more
compounds with $T_c$ as high as 55 K were synthesized by replacing
La with other rare-earth elements such as Sm,
\cite{XHChen1,ZARen1,XHChen2} Ce, \cite{GFChen1} Nd, \cite{ZARen2}
Pr, \cite{ZARen3} and Gd. \cite{HHWen1} Due to their high
superconducting transition temperature, which is second only to the
high-$T_c$ cuprate superconductors, enormous experimental and
theoretical efforts have been put on these materials to clarify
their phase diagram and superconducting mechanism.

These quaternary rare earth transition metal arsenide oxides LnOFeAs
(Ln = La, Sm, Ce, Nd, Pr, and Gd) form tetragonal ZrCuSiAs-type
structure. \cite{Quebe} It is believed that the Fe-As layers are
responsible for the superconductivity and Ln-O layers provide
electron carriers through fluorine doping, or very recently by
simply introducing oxygen vacancies. \cite{ZARen4} Neutron
scattering experiments have demonstrated that the undoped parent
compound LaOFeAs develops long-range SDW-type antiferromagnetic
order below 150 K. \cite{Dai,McGuire} With increasing electron
doping by fluorine, the SDW order is suppressed and
superconductivity emerges, suggesting competing orders in these
systems and similar phase diagram to the one in high-$T_c$ cuprates.
\cite{NLWang1,XHChen2} Theoretically, electron-phonon coupling is
not sufficient to explain superconductivity in LaO$_{1-x}$F$_x$FeAs,
\cite{Boeri} while antiferromagnetic interaction
\cite{ZYLu,YChen,ZYWeng} and Hund's rule ferromagnetic interaction
\cite{XDai,PALee} have been considered as the possible pairing
mechanism.

Among the family of LnO$_{1-x}$F$_x$FeAs, specific heat was only
studied for LaO$_{1-x}$F$_x$FeAs compounds so far.
\cite{HHWen2,RJin1,NLWang1,McGuire} Clear specific heat jump was
observed at the temperature about 150 K for LaOFeAs,
\cite{McGuire,NLWang1} which is accompanied by anomalies on
resistivity, Hall coefficient, and Seebeck coefficient.
\cite{McGuire,NLWang1} Since structural transition was also found at
150 K by neutron scattering \cite{Dai} and X-ray diffraction,
\cite{Nomura} at present it is unclear whether these resistivity and
specific heat anomalies around 150 K are due to the structural or
SDW transition, or to both. For superconducting
LaO$_{0.9}$F$_{0.1-\delta}$FeAs with $T_c \approx$ 28 K,  a
nonlinear magnetic field dependence of the electronic specific heat
coefficient $\gamma(H)$ has been found in the low temperature limit,
which is consistent with the prediction for a nodal superconductor
and suggests an unconventional mechanism for this new
superconductor. \cite{HHWen2} However, it is surprising that no
visible specific heat anomaly was detected near $T_c$ on the raw
data for superconducting LaO$_{1-x}$F$_x$FeAs samples despite the
large Meissner fractions, \cite{HHWen2,RJin1} and only a broadened
anomaly of ($C$(0T)-$C$(9T))/$T$ was observed. \cite{HHWen2} This
result may reflects the low superfluid density in
LaO$_{1-x}$F$_x$FeAs. For comparison, precise specific heat
measurements on other compounds of this family are highly desired.

Here, we systematically study the specific heat of
SmO$_{1-x}$F$_x$FeAs for $0 \leq x \leq 0.2$, with the maximum
$T_c$(onset) = 54 K at $x$ = 0.2. A specific heat jump was observed
at 130 K for undoped $x$ = 0 sample, indicating the structural or
SDW transition. However, this jump disappears in the $x$ = 0.05
sample. The specific heat $C/T$ shows clear anomaly near $T_c$ for
$x$ = 0.15 and 0.20 superconducting samples, which has been absent
in LaO$_{1-x}$F$_x$FeAs. Sharp peak of $C(T)$ appears at 4.6 K for
the parent compound SmOFeAs, and it shifts to 3.7 K with doping
electrons in $x$ = 0.15 sample. This specific heat peak should come
from the antiferromagnetic ordering of Sm$^{3+}$ ions in this
system, as in the electron-doped high-$T_c$ cuprate
Sm$_{2-x}$Ce$_x$CuO$_{4-\delta}$.

The polycrystalline samples with nominal composition
SmO$_{1-x}$F$_x$FeAs ($x$ = 0, 0.05, 0.15, and 0.20) are the same
ones as studied in Ref. 4, synthesized by conventional solid state
reaction. The $x$ = 0 and 0.05 samples are in single phase. A trace
of impurity phases SmOF and SmAs can be observed in $x$ = 0.15
sample, and these impurities are estimated to be less than 10\% in
$x$ = 0.20 sample. Specific heat measurements were performed in a
Quantum Design physical property measurement system (PPMS) via the
relaxation method and the results are presented per mole of atom (J
/ mol K). Magnetic field $H$ = 8 T was applied for the $x$ = 0.15
and 0.20 superconducting samples.

\begin{figure}
\includegraphics[clip,width=8cm]{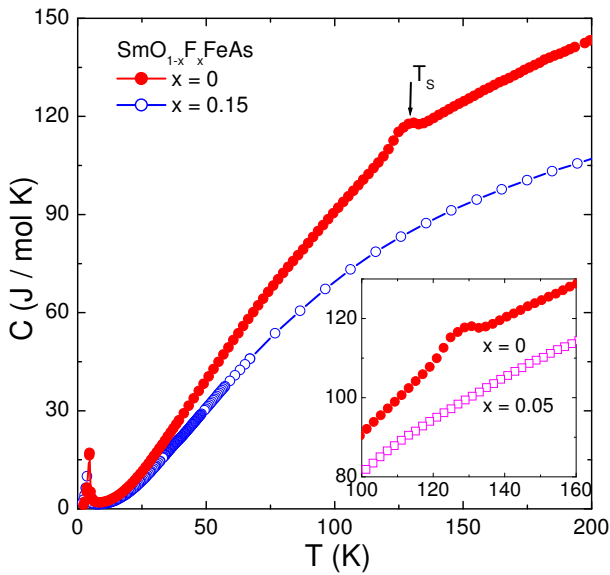}
\caption{(Color online) Specific heat of SmO$_{1-x}$F$_x$FeAs
samples with $x$ = 0 and 0.15. For the $x$ = 0 parent compound,
clear specific heat jump can be seen at 130 K, denoted as the
structural or SDW transition temperature $T_s$. The inset shows the
lack of such jump in the slightly F-doped $x$ = 0.05 sample.}
\end{figure}

The resistivity of this series of SmO$_{1-x}$F$_x$FeAs samples have
been reported previously. \cite{XHChen2} For $x$ = 0 sample, a rapid
resistivity drop below about 130 K was observed. Superconductivity
emerges at $x$ = 0.10, and the $x$ = 0.20 sample has the maximum
superconducting transition temperature $T_c$(onset) = 54 K. The
superconducting volume fractions of the $x$ = 0.20 sample at 5 K
were estimated to be 60\% and 30\% from the susceptibility measured
under zero-field-cool and field-cool conditions at 10 Oe.
\cite{XHChen2}

Fig. 1 shows the specific heat $C(T)$ of SmO$_{1-x}$F$_x$FeAs
samples with $x$ = 0 and 0.15. There is a clear specific heat jump
close to 130 K for $x$ = 0 sample (enlarged in the inset). Similar
jump has been observed in LaOFeAs at 150 K. \cite{McGuire,NLWang1}
The specific heat jump at 130 K of SmOFeAs sample is consistent with
the resistivity drop. \cite{XHChen2} As in LaOFeAs, the specific
heat and resistivity anomalies in SmOFeAs are also attributed to a
structural or SDW transition. We note that the phase transition
temperature $T_s$ in SmOFeAs is about 20 K lower than that in
LaOFeAs. The inset of Fig. 1 shows the lack of specific heat jump in
the slightly F-doped $x$ = 0.05 sample, although there is still a
resistivity drop (but less sharp) at about 110 K. \cite{XHChen2}
This result suggests that electron doping suppresses the magnetic
order and structural distortion in SmOFeAs. Indeed, the static
antiferromagnetic SDW order and structural transition disappear in
the doped superconducting LaO$_{1-x}$F$_x$FeAs samples, shown by
neutron scattering and X-ray diffraction experiments.
\cite{Dai,Nomura} At low temperature there is a sharp peak for both
$x$ = 0 (nonsuperconducting) and 0.15 (superconducting) samples,
which will be discussed later.

\begin{figure}
\includegraphics[clip,width=7cm]{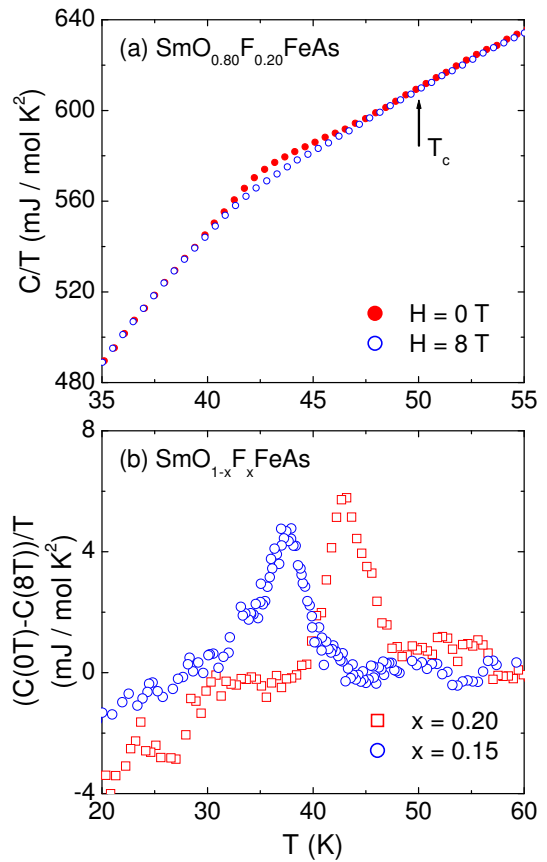}
\caption{(Color online) (a) Specific heat of optimally doped
SmO$_{0.80}$F$_{0.2}$FeAs sample in zero and $H$ = 8 T magnetic
fields, plotted as $C/T$ vs $T$ close to $T_c$. The arrow marks zero
resistivity $T_c$ = 50 K. (b) ($C$(0T)-$C$(8T))/$T$ vs $T$ for the
$x$ = 0.15 and 0.20 samples. One can see a clear specific heat peak
near the zero resistivity $T_c$ for both samples.}
\end{figure}

Fig. 2a plots $C/T$ vs $T$ for the optimally doped
SmO$_{0.80}$F$_{0.2}$FeAs sample in zero and $H$ = 8 T magnetic
fields. Below the zero resistivity $T_c$ = 50 K, one can see a clear
specific heat anomaly. Such anomaly has been absent in the
LaO$_{1-x}$F$_x$FeAs superconducting samples with large Meissner
fractions. \cite{HHWen2,RJin1} The difference may reflect the higher
superfluid density in SmO$_{1-x}$F$_x$FeAs. This is reasonable,
since the maximum $T_c$ of SmO$_{1-x}$F$_x$FeAs is twice that of
LaO$_{1-x}$F$_x$FeAs.

Although 8 T is far away from the upper critical field $H_{c2}$
which is higher than 60 T, \cite{RJin2} we nevertheless plot
($C$(0T)-$C$(8T))/$T$ vs $T$ for the $x$ = 0.15 and 0.20 samples in
Fig. 2b. The obtained specific heat peak near the zero resistivity
$T_c$ is sharper than that in LaO$_{0.9}$F$_{0.1-\delta}$FeAs
sample. \cite{HHWen2}

\begin{figure}
\includegraphics[clip,width=7cm]{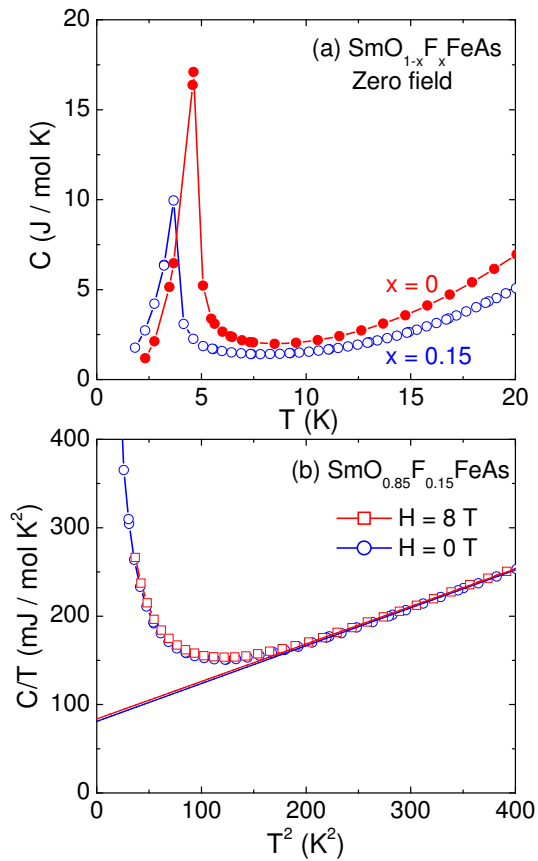}
\caption{(Color online) (a) Low temperature specific heat of
SmO$_{1-x}$F$_x$FeAs samples with $x$ = 0 and 0.15 in zero field.
The sharp peak comes from the antiferromagnetic ordering of Sm$^{3+
}$ ions in this system. (b) $C/T$ vs $T^2$ for the $x$ = 0.15 sample
in $H$ = 0 and 8 T. The lines are linear fits between 14 and 20 K.}
\end{figure}

Below we focus on the low temperature specific heat behavior of
SmO$_{1-x}$F$_x$FeAs at $T < 20$ K. In Fig. 3a, $C(T)$ of the $x$ =
0 sample shows a very sharp peak at 4.6 K. With electron doping in
the $x$ = 0.15 superconducting sample, the peak shifts to 3.7 K and
its height decreases. This low-temperature peak has not been seen in
previous specific heat studies of LaO$_{1-x}$F$_x$FeAs.
\cite{HHWen2,RJin1,NLWang1,McGuire} Since the only difference
between these two materials is the Ln$^{3+}$ ions, i.e. non-magnetic
La$^{3+}$ and magnetic Sm$^{3+}$ ions, this peak may relate to the
magnetic ordering of Sm$^{3+}$ ions. In fact, exactly the same
specific heat behavior at low temperature has been observed in
electron-doped high-$T_c$ cuprate Sm$_{2-x}$Ce$_x$CuO$_{4-\delta}$.
\cite{Ghamaty,Cho} Antiferromagnetic ordering of the Sm$^{3+}$ ions
was found in Sm$_2$CuO$_4$ at $T_N$ = 5.9 K, accompanied by a sharp
specific heat peak. \cite{Ghamaty} By substituting electron donor
element Ce$^{4+}$ for Sm$^{3+}$ ions, the ordering temperature $T_N$
is lowered to 4.7 K in Sm$_{1.85}$Ce$_{0.15}$CuO$_{4-\delta}$ with
$T_c$ = 16.5 K. \cite{Cho} Therefore the sharp specific heat peak
below 5 K in SmO$_{1-x}$F$_x$FeAs manifests the antiferromagnetic
ordering of Sm$^{3+}$ ions. Below $T_N$, the superconductivity
coexists with antiferromagnetism, as in
Sm$_{2-x}$Ce$_x$CuO$_{4-\delta}$.

\begin{figure}
\includegraphics[clip,width=8cm]{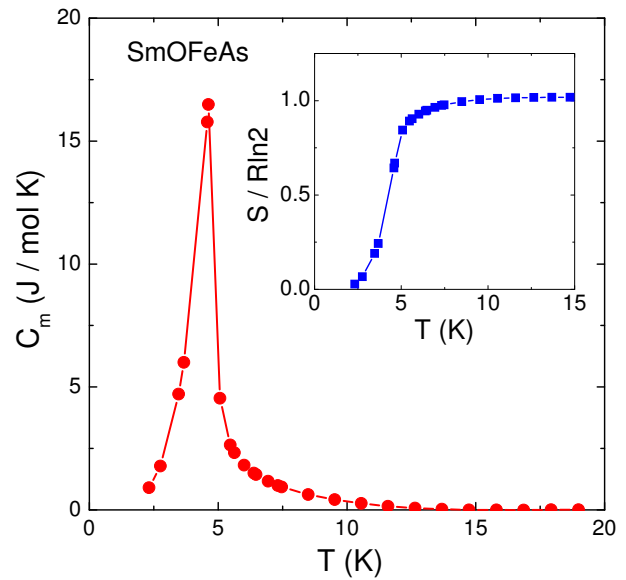}
\caption{(Color online) Magnetic specific heat of SmOFeAs, $C_m = C
- \gamma T - \beta T^3$. Inset: entropy associated with the magnetic
transition.}
\end{figure}

Due to this antiferromagnetic specific heat peak, it is not easy to
extrapolate the electronic specific heat coefficient $\gamma$ in the
low temperature limit. In Fig. 3b, $C/T$ vs $T^2$ is plotted for the
$x$ = 0.15 sample in $H$ = 0 and 8 T. The data between 14 and 20 K
can be linearly fitted by $C/T = \gamma + \beta T^2$, which give
$\gamma$ = 81.0 and 83.7 mJ / mol K$^2$ for $H$ = 0 and 8 T,
respectively. This value of $\gamma$ is much higher than that
obtained in LaO$_{0.89}$F$_{0.11}$FeAs ($T_c \approx$ 28 K),
$\gamma$ = 1.0 mJ / mol K$^2$. \cite{RJin1} For
Sm$_{1.85}$Ce$_{0.15}$CuO$_{4-\delta}$ with $T_c$ = 16.5 K, the
fitting also gave exceptionally large $\gamma$ = 103.2 mJ / mol
K$^2$. \cite{Cho} It was speculated that the effects of magnetic
correlation exist well above $T_N$, thereby making accurate
determination of $\gamma$ difficult. In addition, since the fitting
in Fig. 3b was done at relatively high temperature and over a small
range from 14 to 20 K, the slope $\beta$ may not represent the
phonon specific heat in the low-temperature limit, where it is
proportional to $T^3$. Therefore the resulting large $\gamma$ may be
not reliable. In Fig. 3b, $\gamma$ only increases very slightly in
$H$ = 8 T and we are unable to examine its field dependence for
SmO$_{0.85}$F$_{0.15}$FeAs sample. For superconducting
LaO$_{1-x}$F$_x$FeAs, there is no such antiferromagnetic specific
heat peak and the data were fitted at low temperature, thus the
extrapolated $\gamma \sim$ 1.0 mJ / mol K$^2$ is more reliable and
shows a steep increase with increasing magnetic field.
\cite{RJin1,HHWen2}

Fig. 4 plots the magnetic specific heat of SmOFeAs, $C_m = C -
\gamma T - \beta T^3$, with $\gamma$ = 119.4 mJ / mol K$^2$ and
$\beta$ = 0.56 mJ / mol K$^4$ obtained from the same fitting process
as in Fig. 3b. The magnetic entropy $S$ associated with the magnetic
transition is also calculated from the $C_m(T)$ and shown in the
inset of Fig. 4. With increasing temperature, the entropy rapidly
increases and then saturates to the value of $R$ln2 within
experimental error, indicating that the Sm$^{3+}$ ground state in
the crystal field is a doublet for SmOFeAs.

In summary, we have systematically studied the specific heat of new
iron-based high-$T_c$ superconductor SmO$_{1-x}$F$_x$FeAs. First, a
specific heat jump was observed at 130 K for the undoped $x$ = 0
sample, corresponding to the structural or SDW transition. However,
this jump disappears in the slightly F-doped $x$ = 0.05 sample,
indicating the suppression of the SDW order and structural
distortion by electron doping in this system. Second, a clear
specific heat anomaly can be seen near $T_c$ for superconducting
SmO$_{1-x}$F$_x$FeAs samples, while it has been absent in
LaO$_{1-x}$F$_x$FeAs. This result suggests higher superfluid density
in SmO$_{1-x}$F$_x$FeAs. Finally, sharp specific heat peak shows up
at 4.6 K for $x$ = 0 sample, and it shifts to 3.7 K upon electron
doping in $x$ = 0.15 sample. By comparing with the electron-doped
high-$T_c$ cuprate Sm$_{2-x}$Ce$_x$CuO$_{4-\delta}$, this sharp peak
is attributed to the antiferromagnetic ordering of Sm$^{3+}$ ions in
SmO$_{1-x}$F$_x$FeAs system.

We thank D. L. Feng and Y. Chen for useful discussions. This work is
supported by the Natural Science Foundation of China, the Ministry
of Science and Technology of China (973 project No:
2006CB601001), and National Basic Research Program of China (2006CB922005).\\

$^*$ Electronic address: shiyan$\_$li@fudan.edu.cn


\begin{thebibliography}{99}

\bibitem{Kamihara} Yoichi Kamihara, Takumi Watanabe, Masahiro Hirano, and Hideo Hosono, J. Am. Chem. Soc. {\bf
130}, 3296 (2008).
\bibitem{XHChen1} X. H. Chen, T. Wu, G. Wu, R. H. Liu, H. Chen, D. F. Fang, arXiv:0803.3603.
\bibitem{ZARen1} Zhi-An Ren, Wei Lu, Jie Yang, Wei Yi, Xiao-Li Shen, Zheng-Cai Li, Guang-Can Che, Xiao-Li Dong, Li-Ling Sun, Fang Zhou, Zhong-Xian Zhao, arXiv:0804.2053.
\bibitem{XHChen2} R. H. Liu, G. Wu, T. Wu, D. F. Fang, H. Chen, S. Y. Li, K. Liu, Y. L. Xie, X. F. Wang, R. L. Yang, C. He, D. L. Feng, X. H. Chen, arXiv:0804.2105.
\bibitem{GFChen1} G. F. Chen, Z. Li, D. Wu, G. Li, W. Z. Hu, J. Dong, P. Zheng, J. L. Luo, N. L. Wang, arXiv:0803.3790.
\bibitem{ZARen2} Zhi-An Ren, Jie Yang, Wei Lu, Wei Yi, Xiao-Li Shen, Zheng-Cai Li, Guang-Can Che, Xiao-Li Dong, Li-Ling Sun, Fang Zhou, Zhong-Xian Zhao, arXiv:0803.4234.
\bibitem{ZARen3} Zhi-An Ren, Jie Yang, Wei Lu, Wei Yi, Guang-Can Che, Xiao-Li Dong, Li-Ling Sun, Zhong-Xian Zhao, arXiv:0803.4283.
\bibitem{HHWen1} Peng Cheng, Lei Fang, Huan Yang, Xiyu Zhu, Gang Mu, Huiqian Luo, Zhaosheng Wang, Hai-Hu Wen, arXiv:0804.0835.
\bibitem{Quebe} P. Quebe, T. J. Terbuchte, W. Jeitschko, J. Alloys and Compounds {\bf 302}, 70 (2000).
\bibitem{ZARen4} Zhi-An Ren, Guang-Can Che, Xiao-Li Dong, Jie Yang, Wei Lu, Wei Yi, Xiao-Li Shen, Zheng-Cai Li, Li-Ling Sun, Fang Zhou, Zhong-Xian Zhao, arXiv:0804.2582.
\bibitem{Dai} Clarina de la Cruz, Q. Huang, J. W. Lynn, Jiying Li, W. Ratcliff II, J. L. Zarestky, H. A. Mook, G. F. Chen, J. L. Luo, N. L. Wang, Pengcheng Dai, arXiv:0804.0795.
\bibitem{McGuire} M. A. McGuire, A. D. Christianson, A. S. Sefat, R. Jin, E. A. Payzant, B. C. Sales, M. D. Lumsden, D. Mandrus, arXiv:0804.0796.
\bibitem{NLWang1} J. Dong, H. J. Zhang, G. Xu, Z. Li, G. Li, W. Z. Hu, D. Wu, G. F. Chen, X. Dai, J. L. Luo, Z. Fang, N. L. Wang, arXiv:0803.3426.
\bibitem{Boeri} L. Boeri, O. V. Dolgov, and A. A. Golubov, arXiv:0803.2703.
\bibitem{ZYLu} Fengjie Ma, Zhong-Yi Lu, and Tao Xiang, arXiv:0804.3370.
\bibitem{YChen} Q. Han, Y. Chen and Z. D. Wang, Europhys. Lett. {\bf
82}, 37007 (2008).
\bibitem{ZYWeng} Zheng-Yu Weng, arXiv:0804.3228.
\bibitem{XDai} Xi Dai, Zhong Fang, Yi Zhou and Fu-chun Zhang, arXiv:0803.3982.
\bibitem{PALee} Patrick A. Lee and Xiao-Gang Wen, arXiv:0804.1739.
\bibitem{HHWen2} Gang Mu, Xiyu Zhu, Lei Fang, Lei Shan, Cong Ren, Hai-Hu Wen, arXiv:0803.0928.
\bibitem{RJin1} Athena S. Sefat, Michael A. McGuire, Brian C. Sales, Rongying Jin, Jane Y. Howe, David Mandrus, arXiv:0803.2528.
\bibitem{Nomura} Takatoshi Nomura, Sung Wng Kim, Yoichi Kamihara, Masahiro Hirano, Peter V. Sushko, Kenichi Kato, Masaki Takata, Alexander L. Shluger, and
Hideo Hosono, arXiv:0804.3569.
\bibitem{RJin2} F. Hunte, J. Jaroszynski, A. Gurevich, D.C. Larbalestier, R. Jin, A.S. Sefat, M.A. McGuire, B.C. Sales, D.K. Christen, D. Mandrus, arXiv:0804.0485.
\bibitem{Ghamaty} S. Ghamaty, B. W. Lee, J. T. Markert, E. A. Early, T. Bjornholm, C. L. Seaman, and M. B. Maple, Physica C {\bf 160}, 217 (1989).
\bibitem{Cho} B. K. Cho, Jae Hoon Kim, Young Jin Kim, Beom-hoan O, J. S. Kim, and G. R.
Stewart, Phys. Rev. B {\bf 63}, 214504 (2001).


\end{thebibliography}
\end{document}